\documentstyle[12pt]{article}

\author{Hao Chen\\
Department of Mathematics\\
Zhongshan University\\
Guangzhou,Guangdong 510275\\
People's Republic of China\\
and\\
Department of Computer Science\\
National University of Singapore\\
Singapore 117543\\
Republic of Singapore}
\title{Random low rank mixed states are highly entangled}
\date{October,2001}

\begin{document}

\maketitle
\begin{abstract}
We prove that for many low ranks $r \leq 2m-3$, random rank $r$ mixed states in $H_A^m \otimes H_B^m$ have realtively high Schmidt numbers based on algebraic-geometric separability criterion proved in [1]. This also means that algebraic-geometric separability criterion can be used to detect all low rank entagled mixed states outside a measure zero set.

\end{abstract}

Quantum entanglement was first noted as a feature of quantum mechanics in the famous Einstein, Podolsky and Rosen [2] and Schrodinger [3] papers. Its importance lies not only in philosophical considerations of the nature of quantum theory, but also in applications where it has emerged recently that quantum entanglement is the key ingredient in quantum computation [4] and communication [5] and plays an important role in cryptography [6,7].\\

A mixed state $\rho$ in the bipartite quantum system $H=H_A^m \otimes H_B^n$ is called separable if it can be written in the form $\rho=\Sigma_j p_j |\psi_j><\psi_j| \otimes |\phi_j><\phi_j|$, where $p_j >0$ and $|\psi_j>,|\phi_j>$ are pure states in $H_A^m,H_B^n$. Otherwise it is called entangled, ie., it cannot be prepared by A and B separately. It is realized that entangled states are very important resources in quantum communication, quantum cryptography and quantum computation. Thus one of the fundamental and natural questions concerning quantum entanglement is to estimate how many entamgled or separable states exist among all quantum states ([8],[9],[10],[11]). Similar problems are also considered for continuous variable quantum entanglement ([12],[13]). The main previously known results are: \\

(1) The volumes of separable ,entangled, bound entangled mixed states are not zero ([8]);\\

(2) All mixed states in a neighborhood (this neighborhood can be explicitlly determined ) of the  maximally mixed state are separable ([8],[9],[10]); \\

(3) The volume of separable mixed states approaches zero when dimension goes to infinity ([8]). \\

(4) It is also noted from numerical simulation that the purer the quantum state is, the smaller its possibility of being separable ([8]).\\

There are also applications of the result (2) in analysis of NMR quantum computation ([9]).\\

In [14], Schmidt number of mixed states was introduced as the minimum Schmidt rank of pure states that are needed to construct such mixed states. For a bipartite mixed state $\rho$, it has Schmidt number $k$ if and only if for any decomposition $\rho=\Sigma_i p_i |v_i><v_i|$ for positive real numbers $p_i$'s and pure states $|v_i>$'s, at least one of the pure states $|v_i>$'s has Schmidt rank at least $k$, and there exists such a decomposition with all pure states $|v_i>$'s Schmidt rank at most $k$. It is proved that Schmidt number is entanglement monotone, ie., they cannot increase under local quantum operations and classical communication in [14]. So we can naturally think Schmidt numbers of mixed states as a measure of their entanglement. We say mixed states are highly entangled if their Schmidt numbers are high.\\

The main result of this letter is the following result.\\

{\bf Theorem 1.} {\em There exists a subset $Z(r)$ defined by algebraic equations in the space $M(r)$ of all rank $r$ mixed states in bipartite quantum system $H_A^m \otimes H_B^m$ , such that, Schmidt numbers of mixed states in $M(r)\setminus Z(r)$ are at least\\
1) $[\sqrt{m}]-1$ if $r\leq m$;\\
2) $\frac{m}{r-m+[\sqrt{m}]+1}$ if $r>m$;\\
3) $3$ if $\frac{3m}{2}-5 \geq r >m \geq 169$;\\
4) $2$ if $r \leq 2m-3$.}\\

Here $[x]$ of a positive real number $x$ means the integral part of $x$.\\

From now on, ``generic'' elememts of a set means the elements of this set outside an algebraic set (ie. zero locus of some algebraic equations, see [13]).\\

We should note that the algebraic set $Z(r)$ ([21]) defined as zero locus of algebraic equations has volume zero under any reasonable measure. Thus it is known from Theorem 1 that random low rank mixed states in $H_A^m \otimes H_B^m$ are highly entangled, for example, random rank $r \leq m+\sqrt{m}$ mixed states have their Schmidt numbers at least $[\sqrt{m}]/2-1$ from 1) and 2) of Theorem 1.\\

For example, for rank 1 mixed states (pure states) $\phi=\Sigma_{i,j}^m a_{ij} |ij>$ it is well-known that the Schmidt number of $\phi$ is just the rank of the matrix $A=(a_{ij})_{1\leq i,j \leq m}$. Thus the pure states outside the algebraic set defined by $det A=0$ have Schmidt number $m$, thus highly entagled. This is previously known result ([15],[8]).\\

Our special attention to low rank mixed states is partially motivated from previous works in [16],[17],[18], where  it is proved that PPT low rank $r \leq n$  mixed states in $H_A^m \otimes H_B^n$ ($m\leq n$) are separable and a constructive method to check separability of low rank mixed states is presented. Their results suggested that in some sence there are not too ``many'' low rank separable mixed states.\\

After the posting of 1st version of this paper, 4) in Theorem 1 is extened to many other ``low `` ranks and arbitrary multipartite cases in [19]. The proof there is based on a dimension-counting argument and Horodecki range criterion [20]. \\

Our result Theorem 1 is proved by a lower bound of Schmidt numbers of bipartite mixed states which is derived from the emptyness of certain algebraic-geometric invariants of these mixed states developed in [1] (Theorem 2 below). For any concrete bipartite mixed states given explicitly, it is realtively easy to calculate these algebraic-geometric invarinants and then get strong lower bounds of their Schmidt numbers (see Example 1,2 and  Corollary 1,2 below). Thus our method is constructive.\\

For bipartite mixed states $\rho$ in $H=H_A^m \otimes H_B^n$, algebriac sets $V_A^k(\rho)$ in $CP^{m-1}$ (respectively $V_B^k(\rho)$ in $CP^{n-1}$) are introduced in [1] as the {\em degenerating locus} of the measurement of the mixed states by separable pure states. They are non-local invariants of $\rho$, ie., they are invariant when local unitary transformations applied to the mixed states. Moreover these algebraic sets are independent of the eigenvalues and only measure the positions of the eigenvectors of the mixed states. \\

Let us recall the basic facts about algebraic-geometric invariants of bipartite mixed states in [1]. For any bipartite mixed states $\rho$ on $H_A^m \otimes H_B^n$ , we want to understand it by measuring it with separable pure states, ie., we consider the $<\phi_1 \otimes \phi_2 |\rho|\phi_1 \otimes \phi_2>$ for any pure states $\phi_1 \in H_A^m$ and $\phi_2 \in H_B^n$. For any fixed $\phi_1 \in P(H_A^m)$, where $P(H_A^m)$ is the projective space of all pure states in $H_A^m$,  $<\phi_1 \otimes \phi_2 |\rho|\phi_1 \otimes \phi_2>$ is a Hermitian bilinear form on $H_B^n$, denoted by $<\phi_1|\rho|\phi_1>$ . We consider the {\em degenerating locus } of this bilinear form, ie., $V_A^k(\rho)=\{\phi_1 \in P(H_A^m): rank (<\phi_1|\rho|\phi_1>) \leq k\}$ for $k=0,1,...,n-1$. We can use the coordinate form of this formalism. Let $\{|11>,...,|1n>,...,|m1>,...,|mn>\}$ be the standard orthogonal base of $h_A^m \otimes H_B^n$ and $\rho$ be an arbitrary mixed states. We represent the matrix of $\rho$ in the base $\{|11>,...|1n>,...,|m1>,...,|mn>\}$, and consider $\rho$ as a blocked matrix $\rho=(\rho_{ij})_{1 \leq i \leq m, 1 \leq j \leq m}$ with each block $\rho_{ij}$ a $n \times n$ matrix corresponding to the $|i1>,...,|in>$ rows and the $|j1>,...,|jn>$ columns. For any pure state $\phi_1=r_1|1>+...+r_m|m> \in P(H_A^m)$ the matrix of the Hermitian linear form $<\phi_1|\rho|\phi_1>$ with the base $|1>,...,|n>$ is $\Sigma_{i,j} r_ir_j^{*} \rho_{ij}$. Thus the ``degenerating locus'' is actually as follows.\\

$$
\begin{array}{ccccc}
V_{A}^k(\rho)=\{(r_1,...,r_m)\in CP^{m-1}:rank( \Sigma_{i,j}r_ir_j^{*} \rho_{ij}) \leq k\}
\end{array}
$$
for $k=0,1,...,n-1$. Similarly $V_{B}^k (\rho) \subseteq CP^{n-1}$ can be defined. Here * means the conjugate of complex numbers. It is known from Theorem 1 and 2 of [1] that these sets are algebraic sets (zero locus of several multi-variable polynomials, see [21]) and they are invariants under local unitary transformations depending only on the eigenvectors of $\rho$. Actually these algebraic sets can be computed easily as follows.\\

Let $\{|11>,...,|1n>,...,|m1>,...,|mn>\}$ be the standard orthogonal base of $H_A^m \otimes H_B^n$ as above and $\rho= \Sigma_{l=1}^{t} p_l |v_l><v_l|$ be any given  representation of $\rho$ as a convex combination of projections with $p_1,...,p_t >0$. Suppose $v_l=\Sigma_{i,j=1}^{m,n} a_{ijl} |ij>$ , $A=(a_{ijl})_{1\leq i \leq m, 1 \leq j \leq n, 1 \leq l \leq t}$ is the $mn \times t$ matrix. Then it is clear that the matrix representation of $\rho$ with the base $\{|11>,...,|1n>,...,|m1>,...,|mn>\}$ is $AP(A^{*})^{\tau}$, where $P$ is the diagonal matrix with diagonal entries $p_1,...,p_t$. We may consider the $mn\times t$ matrix $A$ as a $m\times 1$ blocked matrix with each block $A_w$, where $w=1,...,m$, a $n\times t$ matrix corresponding to $\{|w1>,...,|wn>\}$. Then $V_A^k(\rho)$ is just the algebraic set in $CP^{m-1}$ as the zero locus of the determinants of all $(k+1) \times (k+1)$ submatrices of $\Sigma_i^m r_i A_i$.\\

The following obsevation is the the key point of the proof of Theorem 1. From Lemma 1 in [20], the range of $\rho$ is the linear span of vectors $|v_1>,...,|v_t>$. We take any $dim(range(\rho))$ linear independent vectors in the set $\{|v_1>,...,|v_t>\}$, say they are $|v_1>,...,|v_s>$ , where $s=dim(range(\rho))$. Let $B$ be the $mn \times s$ matrix with columns corresponding to the $s$ vectors $|v_1>,...,|v_s>$'s coordinates in the standard base of $H_A^m \otimes H_B^n$. We consider $B$ as $m \times 1$ blocked matrix with blocks $B_1,...,B_m$ $n \times s$ matrix as above. It is clear that $V_A^k(\rho)$ is just the zero locus of determinants of all $(k+1) \times (k+1)$ submatrices of $\Sigma_i^m r_iB_i$, since any column in $\Sigma_i r_i A_i$ is a linear combination of columns in $\Sigma_i r_i B_i$.\\

We have the following result, which gives a strong lower bound of Schmidt numbers of bipartite mixed states and is the key of the proof of Theorem 1.\\

{\bf Theorem 2.} {\em Let $\rho$ be a mixed state on $H_A^m \otimes H_B^m$ of rank $r$ and Schmidt number $k$. Suppose $V_A^{m-t}(\rho)=\emptyset$, then $k \geq \frac{m}{r-m+t}$.}\\

{\bf Proof.} Take a representation $\rho= \Sigma_{i=1}^t p_i |v_i><v_i|$ with $p_i$'s positive,  and the maximal Schmidt rank of $v_i$'s is $k$. As observed above, it only need to take $r$ linear independent vectors in $\{v_1,...,v_t\}$ to compute the rank of $\Sigma_i r_i A_i$. For the purpose that the rank of these $r$ columns in $\Sigma_i r_i A_i$ is not bigger than $m-t$, we just need $r-m+t$ of these columns are zero. On the other hand, from Proposition 1 in [1], the dimension of the linear subspace $(r_1,...,r_m) \in H_A^m$, such that the corresponding column of $v_i$ in $\Sigma_i r_i A_i$ is zero,  is exactly $m-k(v_i)$ where $k(v_i)$ is the Schmidt rank of $v_i$. Thus we know that there is at least one nonzero $(r_1,...,r_m)$ such that $\Sigma_i r_i A_i$ is of rank smaller than $m-t+1$ if $ m >k(r-m+t)$. The conclusion is proved.\\

Theorem 2 can be used to give strong results for Schmidt numbers of mixed states as illustrated in the following example.\\

{\bf Example 1.} Let $\rho$ be a rank 2 mixed state in $H_A^m \otimes H_B^m$ of the form $\rho=\lambda_1 |v_1><v_1|+\lambda_2 |v_2><v_2|$, where $\lambda_1$ and $\lambda_2$ positive, $v_1$ and $v_2$ are linear independent unit vectors and of the form $v_1=\Sigma_{ij}a_{ij}^1|ij>$ and $v_2=\Sigma_{ij} a_{ij}^2 |ij>$. Suppose the linear span by the $2m$ rows of the matrices $A^1=(a_{ij}^1)_{1\leq i,j\leq m}$ and $A^2=(a_{ij}^2)_{1\leq i,j \leq m}$ is of dimension $m$ (We should note that this is a condition satified by generic rank two mixed states, since for all rank 2 mixed states outside an algebraic set, the $2m \times m$ matrix consisting of $2m$ rows of $A^1$ and $A^2$ has rank $m$). Then Schmidt number of $\rho$ is at least $\frac{m}{2}$.\\ 

For the proof, we just need to take $r=2,t=m$ in Theorem 2. From the condition that the $2m \times m$ matrix consiting of $2m$ rows of $A^1$ and $A^2$ has rank $m$, we can easily get $V_A^0(\rho)=\emptyset$. Thus from Theorem 2, we get the conclusion.\\ 

For example, the following rank 2 mixed states $\rho_{\lambda_1,\lambda_2}=\frac{1}{\lambda_1+\lambda_2}(\lambda_1 |v_1><v_2|+ \lambda_2 |v_2><v_2|)$ in bipartite quantum system $H_A^5 \otimes H_B^5$, where $\lambda_1$ and $\lambda_2$ are real positive real numbers and \\

$$
\begin{array}{cccccc}
v_1=\frac{1}{\sqrt{2}}(|11>+|22>)\\
v_2=\frac{1}{2}(|33>+|44>+|55>+|45>)\\
\end{array}
$$

have their Schmidt numbers at least 3.\\

Actually example 1 can be generalized as follows.\\

{\bf Corollary 1.} {\em Let $\rho=\Sigma_{t=1}^r p_i |v_t><v_t|$ be a rank $r < m$ mixed state in $H_A^m \otimes H_B^m$, where $p_1,...,p_r$ are positive numbers and $v_t=\Sigma_{ij} a_{ij}^t|ij>, A^t=(a_{ij}^t)_{1 \leq i,j \leq m}$. Suppose that the linear span of all $rm$ rows of matrices $A^1,...,A^r$ is of dimension $m$. Then Schmidt number of $\rho$ is at least $\frac{m}{r}$,thus entangled.}\\

{\bf Proof.} We take $t=m$ in Theorem 2 and from the condition about matrices $A^1,...,A^r$ it is clear $V_A^0(\rho)=\emptyset$. Thus the conclusion follows from Theorem 2.\\

Corollay 1 implies that if a mixed state is mixed by not too many pure states and one of these pure states has highest Schmidt rank, then the mixed state has a relatively high Schmidt number. It is clear that the condition of Corollary 1 is satisfied by generic rank $r<m$ mixed states.\\

{\bf Example 2.} Let $\rho_{\lambda_1,\lambda_2,\lambda_3}=\frac{1}{\lambda_1+\lambda_2+\lambda_3}(\lambda_1 |v_1><v_2|+\lambda_2 |v_2><v_2|+\lambda_3 |v_3><v_3|)$ in bipartite quantum system $H_A^7 \otimes H_B^7$, where $\lambda$'s  are positive real numbers, $v_1=\frac{1}{\sqrt{7}}(|11>+\cdots+|77>)$ and $v_2$ and $v_3$ are arbitray pure states. Then Schmidt numbers of $\rho_{\lambda_1,\lambda_2,\lambda_3}$ are at least 3.\\

Generally the follwing result is valid.\\

{\bf Corollary 2.}{\em Let $\rho=\Sigma_{i=1}^r p_i |v_i><v_i|$ be a mixed state in $H_A^m \otimes H_B^m$, where $p_1,...,p_r$ are positive real numbers and Schmidt rank of $v_1$ is $m$. Then Schmidt number of $\rho$ is at least $\frac{m}{r}$ and thus\\
1) Schmidt number of $\rho$ is at least 3 if $r <\frac{m}{2}$;\\
2) $\rho$ is entangled when $r<m$.}\\

From Example 1,2 and Corollary 1,2 we can see that our method is constructive to check Schmidt numbers of mixed states by just calculating rank of some numerical matrices.\\

For the purpose to prove Theorem 1, we need to recall a well-known result in the theory of determinantal varieties (see Proposition in p.67 of [22]). Let $M(m,n)=\{(x_{ij}): 1\leq i \leq m, 1 \leq j \leq n\}$ (isomorphic to $CP^{mn-1}$) be the projective space of all $ m \times n$ matrices. For a integer $0 \leq k \leq min\{m,n\}$, $M(m,n)_k$ is defined as the locus $\{A=(x_{ij}) \in M(m,n): rank(A) \leq k\}$. $M(m,n)_k$ is called generic determinantal varieties.\\

{\bf Proposition 1.} {\em $M(m,n)_k$ is an irreducible algebriac subvariety of $M(m,n)$ of codimension $(m-k)(n-k)$.}\\

From Theorem 2 and Proposition 1 the following result can be proved.\\

{\bf Corollary 3.} {\em Generic rank $m$ mixed states in $H_A^m \otimes H_B^m$ have Schmidt numbers at least $[\sqrt{m}]-1$.}\\

{\bf Proof.} We take $r=m,t=[\sqrt{m}]+1$. For rank $m$ mixed state $\rho$ we take its spectral decomposition $\rho=\Sigma_i^m p_i |\phi_i><\phi_i|$, from Theorem 2 and its proof in [1], $V_A^{m-t}(\rho)$ can be computed from this spectral decomposition. Thus $V_A^{m-t}(\rho)$ is the locus of the condition that the $m \times m$ matrix corresponding to eigenvectors $\phi_1,...,\phi_m$ (as described in [1]) has rank smaller than $m-t+1$. From proposition 1, for generic rank $m$ mixed states $\rho$ (ie., generic $m\times m$ matrices), $V_A^{m-t}(\rho)$ in $CP^{m-1}$ has codimension $t^2 >m-1$, thus empty. From Theorem 2 the conclusion is proved.\\

Now we can prove Theorem 1, the idea is basically the same as the proof of Corollary 1,2,3, ie., we take suitable $t$ such that $V_A^{m-t}(\rho)$ has codimension larger than $m-1$ and then apply Theorem 2 to get the conclusion.\\

{\bf Proof of Theorem 1.} Similarly as the argument for Corollary 3, for any given $m,r$ for the purpose that $V_A^{m-t}(\rho)=\emptyset$ for generic mixed states in $H_A^m \otimes H_B^m$ we just need $(m-(m-t))(r-(m-t)) \geq m$ from Proposition 1. \\

We take $t=m-r+\sqrt{m}+1$ in case 1) and $t=\sqrt{m}+1$ in case 2) , the conclusions of 1) and 2) are proved.\\

It is clear that random $r \leq \frac{4m-3-3[\sqrt{m}]}{3}$ has their Schmidt numbers at least $3$ from 1) and 2). For ranks $\frac{4m-3-3[\sqrt{m}]}{3} \leq r \leq \frac{3m}{2}-5$ we can take $t=4$, the conclusion of 3) is proved.\\

For mixed states of rank $r=2m-3$, we take $t=2$. A similar argument as above implies that $V_A^{m-2}(\rho)$ in $CP^{m-1}$ has codimension $2(m-1)>m-1$, thus empty for generic $2m-3$ mixed states. Hence from Theorem 2, Schmidt numbers of generic rank $2m-3$ mixed states are at least $\frac{m}{2m-3-m+2}=\frac{m}{m-1}$. For other ranks we can use a similar argument to get the conclusion.\\

Since algebraic-geometric invariants $V_A^k(\rho)$ of mixed states $\rho$ are only dependent on range of $\rho$. Thus our results actually imply that generic low dimension $r \leq 2m-3$ subspaces of $H_A^m \otimes H_B^m$ cannot be a linear span of separable pure states(recall range criterion of P. Horodecki in [20]). However this is not true for high dimensional subspaces of $H_A^m \otimes H_B^m$. In the following example it is proved that generic dimension 3 subspaces of $H_A^2 \otimes H_B^2$ are linear span of separable pure states.\\

{\bf Example 3.} For any given 4 complex numbers $a,b,c,d$ satisfying that $d\neq 0 $ and $ad \neq bc$, it is easy to check that the following $3$ product vectors $v_1,v_2,v_3$ are linear independent and orthogonal to the vector $a|11>+b|12>+c|21>+d|22>$ in $H_A^2 \otimes H_B^2$.\\

$$
\begin{array}{ccccccccccc}
v_1=-c|11>+a|21>\\
v_2=-d|12>+b|22>\\
v_3=-(c+d)|11>-(c+d)|12>+(a+b)|21>+(a+b)|22>\\
\end{array}
$$

We have a family of dimension 3 subspaces span by $v_1,v_2,v_3$ in $H_A^2 \otimes H_B^2$. Since any dimension 3 subspace of $H_A^2 \otimes H_B^2$ has only one normal direction, thus this example showed that generic dimension 3 subspaces (Here generic dimension 3 subspaces means that their normal directions $a|11>+b|12>+c|21>+d|22>$ are outside the algebraic set defined by $d=0$ or $ad=bc$.) of $H_A^2 \otimes H_B^2$ are linear span of seaprable pure states. In high rank mixed state entanglement , it seems people cannot only use ranges to determine whether mixed states are entangled and eigenvalues play certain role, as partially manisfested in result (2) (of [8],[9],[10] mentioned in introduction) for highest rank mixed states.\\

In conclusion, we have proved that for many cases of low ranks $r$ ,random rank $r$ mixed states in $H_A^m \otimes H_B^m$ have relatively high Schmidt numbers and thus highly entangled. The results presented here also indicated that algebraic-geometric separability criterion of [1] is strong enough to detect generic low rank entangled mixed states. Our method can be used constrcutively to get strong lower bounds of Schmidt numbers of low rank bipartite mixed states.\\

The author acknowledges the support from NNSF China, Information Science Division, grant 69972049.\\

e-mail: dcschenh@nus.edu.sg\\

\begin{center}
REFERENCES
\end{center}

1.Hao Chen, quant-ph/01008093\\

2.A.Einstein, B.Podolsky and N.Rosen, Phys. Rev. 47,777(1935)\\

3.E.Schrodinger, Proc.Camb.Philos.Soc.,31,555(1935)\\

4.R.Jozsa, in The Geometric Universe, edited by S.Huggett, L.Mason, K.P.Tod, S.T.Tsou, and N.M.J.Woodhouse (Oxford Univ. Press, 1997)\\

5.A.Ekert, Phys Rev.Lett.67, 661(1991)\\

6.C.H.Bennett, G.Brassard, C.Crepeau, R.Jozsa, A.Peres and W.K.Wootters, Phys.Rev.Lett 70, 1895 (1993)\\

7.C.H.Bennett, G.Brassard, S.Popescu, B.Schumacher, J.Smolin and W.K.Wootters, Phys. Rev.Lett. 76, 722(1996)\\

8.K. Zyczkowski, P.Horodecki, A.Sanpera and M.Lewenstein, Phys. Rev. A., 58:883-892 (1998)\\

9.S.L.Braunstein,C.M.Caves,R.Jozsa,N.Linden,S.Popescu and R.Schack, Phys.Rev.Lett.,83,1054(199)\\

10.G.Vidal and R.Tarrach, Phys. Rev., A 59,141(1999)\\

11.M.Horodecki, P.Horodecki and R.Horodecki, in Quantum Information--Basic concepts and experiments, edited by G.Adler and M.Wiener (Springer Berlin, 2000)\\

12.R.Clifton and H.Halvorson, Phys. Rev. A., 61, 012108(2000)\\

13.P.Horodecki, J.I.Cirac nad M.Lewenstein, quant-ph/0103076, v3\\

14.B.M.Terhal and P.Horodecki, Phys. Rev. A R040301, 61(2000)\\

15.S.Popescu,Phys.Rev.Lett. 72,797(1994),74,2619(1995)\\

16.B.Kiraus, J.I.Cirac, S.Karnas and M.Lewenstein, Phys. Rev. A 61, 062302 (2000)\\

17.P.Horodecki ,M.Lewenstein, G.Vidal and J.I. Cirac, Phys. Rev. A 62, 032310 (2000)\\

18.S.Karnas and M.Lewenstein, Phys. Rev. A, 64 ,042313 (2001)\\

19.R.Lockhart, quant-ph/0111051\\

20.P.Horodecki, Phys.Lett. A 232 333(1997)\\

21.J.Harris, Algebraic Geometry, A First course, GTM 133, Springer-Verlag 1992\\

22.E.Arbarello, M.Cornalba, P.A.Griffiths and J.Harris, Geometry of algebraic curves,Volume I, Springer-Verlag, 1985, Chapter II''Determinantal Varieties''\\ 

\end{document}